\documentclass[reprint,prd,twocolumn,superscriptaddress,showpacs,nofootinbib,floatfix,preprintnumbers]{revtex4-1}
\pdfoutput=1
\usepackage{amsmath}
\usepackage{graphicx,bm}
\usepackage{color}
\usepackage{hyperref} 
\hypersetup{linktocpage=true}
\usepackage[all]{hypcap}
\hypersetup{bookmarks=true,         
   unicode=false,          
   pdftoolbar=true,        
   pdfmenubar=true,        
   pdffitwindow=false,     
   pdfstartview={FitH},    
   pdftitle={My title},    
   pdfauthor={Author},     
   pdfsubject={Subject},   
   pdfcreator={Creator},   
   pdfproducer={Producer}, 
   pdfkeywords={keyword1} {key2} {key3}, 
   pdfnewwindow=true,      
   colorlinks=true,       
   linkcolor=blue,          
   citecolor=magenta,        
   filecolor=magenta,      
   urlcolor=cyan           
}

\begin{document}
\preprint{DAMTP-2015-04-13} 
\title{Consistency of the recent ATLAS $Z+E_T^{\rm miss}$ excess in a simplified GGM model}

\author{Ben Allanach}
\affiliation{DAMTP, CMS, Wilberforce Road, University of Cambridge, Cambridge, CB3 0WA, United Kingdom}
\author{Are Raklev}
\affiliation{Department of Physics, University of Oslo, N-0316 Oslo, Norway} 
\author{Anders Kvellestad}
\affiliation{Department of Physics, University of Oslo, N-0316 Oslo, Norway} 

\begin{abstract}
ATLAS recently reported a $3\sigma$ excess in a leptonic-$Z+E_T^{\rm
  miss}$ channel. This was interpreted in the literature in a simplified
General Gauge Mediation model containing a gluino, a higgsino next-to-lightest
supersymmetric particle (NLSP) and a
gravitino lightest supersymmetric particle (LSP). We test the consistency of this explanation in lieu of the
results of the corresponding search in CMS, and other LHC searches for New
Physics. 
Due to non-decoupling effects from squarks the parameter space of these models
is split into two regions; in one region additional leptons via top
quark production is expected, while the other region sees a large probability
for zero-lepton events. After combining the relevant constraints we find that
these models cannot explain the ATLAS excess. 
\end{abstract}
\maketitle

\section{Introduction}
A recent ATLAS search for beyond the standard model physics in a channel with two leptons, consistent with the production of a $Z$-boson, large missing transverse momentum ($E_T^{\rm miss}$), and at least two jets, reports a
3$\sigma$ excess~\cite{Aad:2015wqa} for 20.3 fb$^{-1}$ of integrated
luminosity at a center of mass energy of 8 TeV. The other general purpose LHC
experiment, CMS, has reported on a similar search, also with the full Run-I
data set~\cite{Khachatryan:2015lwa}, seeing no excess. However, the cuts used
in the two searches are different, and the observed ATLAS excess may {\it a
  priori}\/ be consistent with the CMS results. Also, because the cuts are
different, for some particular interpretation in terms of a new physics model one expects the
predicted signal rates in each analysis to depend upon the signal
kinematics. Hence, the relative number of predicted signal events in ATLAS as
compared to CMS will depend in general upon the assumed interpretation, as
well as on its parameters.  

\begin{figure}[h!]
  \includegraphics[width=4cm]{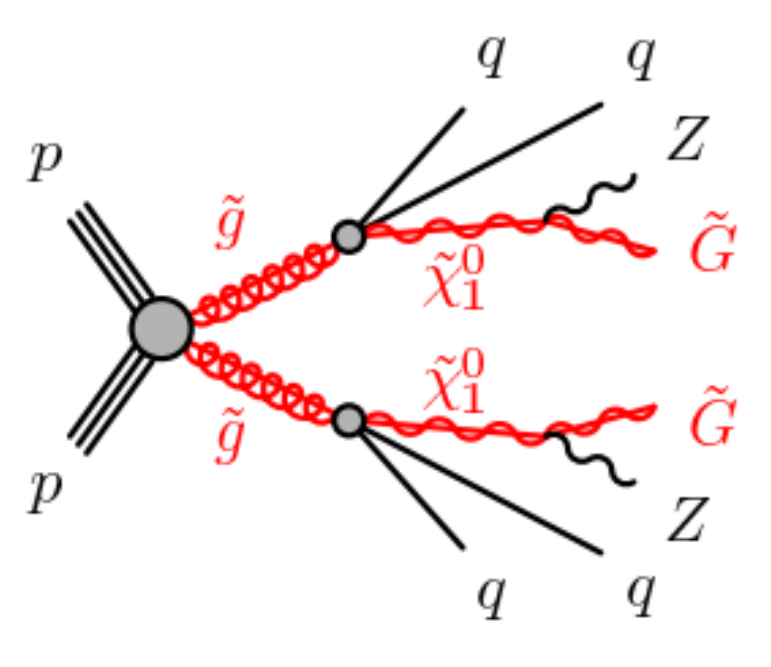}
  \vspace{-0.3cm}
  \caption{Hypothesized GGM decay mode contributing to the ATLAS excess.}
  \label{fig:diag}
\end{figure}

In this article we investigate the consistency of the ATLAS excess with the CMS
analysis, and with other searches at the LHC, for a General Gauge Mediation
(GGM) model with a gravitino LSP~\cite{Meade:2008wd}. The simplified model
used here is inspired by \cite{Ruderman:2011vv}, and contains only three free
parameters: the gaugino mass $M_3$, fixing the gluino mass, $\tan\beta$, the ratio of the two higgs field vacuum expectation values, and $\mu$, the
superpotential parameter, giving the mass of the higgsino NLSP, as well as one
additional neutralino and one chargino, both dominantly higgsino.  
The excess can then be interpreted as stemming from gluino pair production and
the decay chain $\tilde g \to qq \tilde\chi_1^0 \to qq Z \tilde G$, depicted in Fig.~\ref{fig:diag}, with a
leptonic $Z$-decay. In doing so we follow the model chosen by ATLAS to
interpret the results of their analysis. 

As the leptonic branching ratio of the $Z$ is small, this model may also come in to
conflict with recent searches for supersymmetry (SUSY) via jets and missing
energy channels. This was briefly commented on in \cite{Ellwanger:2015hva}, but no detailed analysis of
the parameter space was performed, and the article goes on to interpret the ATLAS
results in an alternative model with $Z$s coming from the decay $\tilde\chi^0_2
\rightarrow Z \tilde\chi^0_1$. The ATLAS excess was also interpreted in
\cite{Barenboim:2015afa}, using two benchmark points in a GGM model with
properties very similar to the model used by ATLAS, and in
\cite{Vignaroli:2015ama}, in the context of a composite Higgs/Randall Sundrum
model, with heavy Kaluza-Klein gluon resonances decaying to vector-like quarks.

\section{Model \label{sec:model}}

For the model used in the ATLAS interpretation the remaining minimal
supersymmetric standard model (MSSM) parameters were set as follows: the gaugino soft masses $M_1=M_2=1$~TeV, the sfermion soft masses $m_{\tilde f}=1.5$~TeV,  and a gravitino mass light enough for the NLSP decays to be prompt. 
With $m_{\tilde q} = 1.5$~TeV, squark--gluino production dominates over gluino
pair production for gluino masses above $\sim 1$~TeV. 
Squark--squark production would dominate sparticle production 
for high enough gluino masses. 
Also, with $M_1=M_2=1$~TeV more complicated decay chains open up for gluinos in this mass range, and the NLSP will no longer be dominantly higgsino for values of $\mu$ close to $1$~TeV.\footnote{Alternatively, MSSM scenarios with a wino (or bino) NLSP could be considered, however, they have $BR(\tilde\chi_1^0 \to \gamma\tilde G)>0.23\,(0.77)$~\cite{Meade:2009qv}, which should be easy to exclude from $\gamma$+$E_T^{\rm miss}$ searches.}

In order to explore a wider range of $\mu$ and $M_3$ values without
introducing such complications in the phenomenology we adopt a simpler model
with $M_1=M_2=1.5$~TeV and sfermions completely decoupled at $m_{\tilde
  f}=4.5$~TeV, keeping in mind that lowering the squark mass scale generally
will lead to stronger bounds on the model. The mass parameters are defined at a scale of $\sqrt{m_{\tilde t_1}m_{\tilde t_2}} \sim 4.5$ TeV. 

The gravitino mass is given by the scale of SUSY breaking, but must be very
light for the NLSP to decay promptly; we set it to be effectively zero for the collider
simulation. All results are presented for $\tan\beta=1.5$ and
$\tan\beta=30$. The choice of low $\tan\beta$ is made in order to maximize the
branching ratio $\tilde\chi_1^0 \to Z \tilde G$, which is in competition with
$\tilde\chi_1^0 \to h \tilde G$, and to a smaller extent $\tilde\chi_1^0 \to \gamma\tilde G$. For
low values of $\tan \beta$, $\mu>0$ and a higgsino NLSP, this is approximately
100\%~\cite{Ambrosanio:1996jn,Dimopoulos:1996yq,Meade:2009qv}. Increasing
$\tan\beta$ will decrease the signal. The two values used thus explore
different parts of the parameter space. 

The lightest higgs mass is simply set to the experimentally measured value  of $m_h=125.09\pm0.24$~GeV~\cite{Aad:2015zhl} by assuming extra operators in the higgs sector, {\it e.g.}\ by using dimension-5 operators as proposed in~\cite{Dine:2007xi}. 

\begin{figure}[h!]
  \includegraphics[width=8cm]{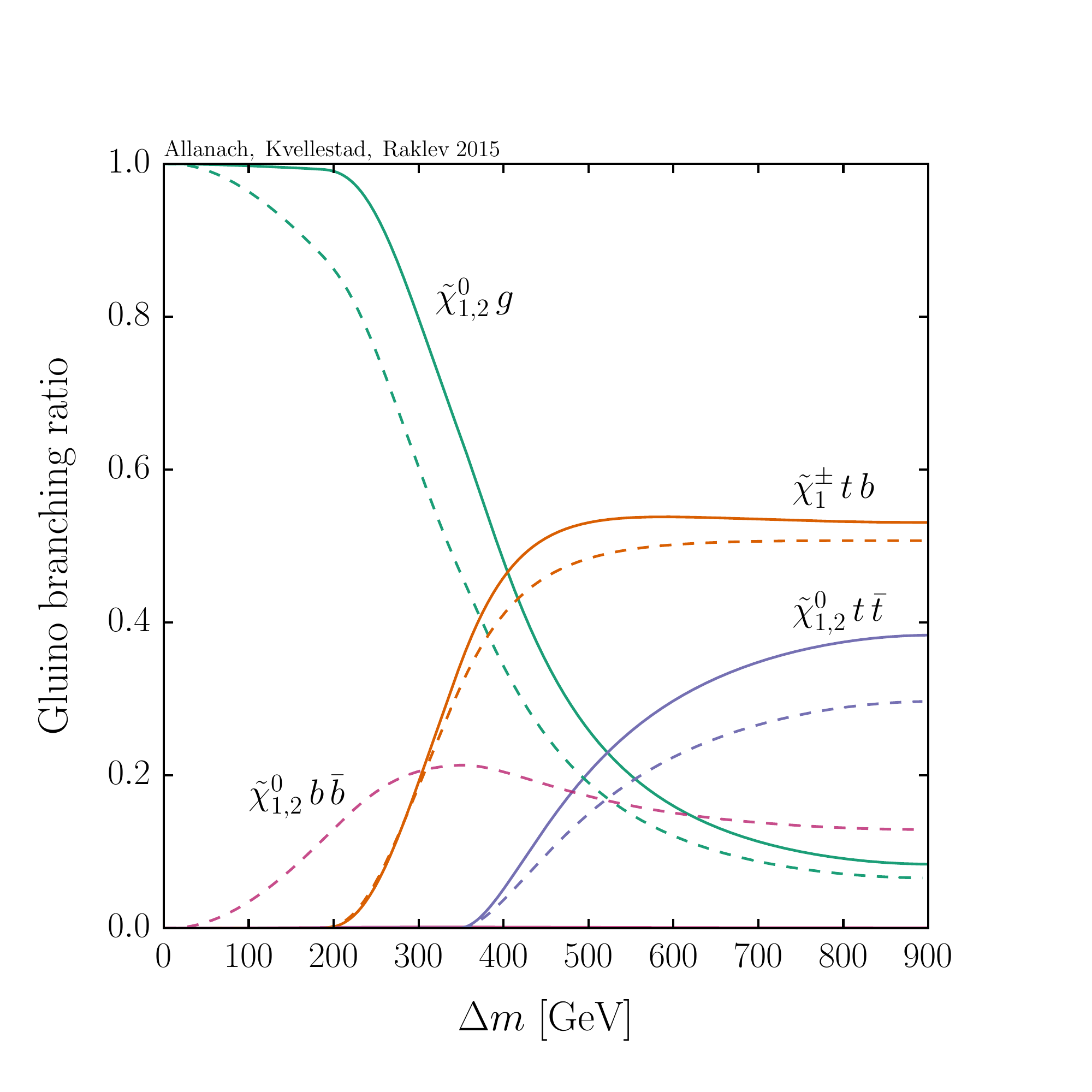}
  \vspace{-0.3cm}
  \caption{Branching ratios for the gluino as a function of the
    gluino--neutralino mass difference with $\tan\beta = 1.5$ (solid lines)
    $\tan\beta = 30$ (dashed lines). The gluino mass is fixed at $m_{\tilde g}
    = 900$ GeV. Some lines are not visible because they
  are at very low values of the gluino branching ratio.}
  \label{fig:gBR}
\end{figure}

The relative squark masses (and to a smaller extent the value of $M_1$ and
$M_2$) determine the branching ratio of the gluino decay into the various
quark flavours. Since the NLSP is dominantly higgsino there will necessarily
be large branching ratios into third generation quarks. When these are
kinematically forbidden, the loop induced decays $\tilde g \to g
\tilde\chi_{1,2}^0$ 
become important. 
Due to the importance of decays involving third generation quarks, $\tan\beta$ also affects the gluino branching ratios. Further complicating matters is the existence of multiple higgsinos at roughly the same mass ($\tilde\chi_1^0$, $\tilde\chi_2^0$ and $\tilde\chi_1^\pm$)\footnote{We note that a co-NLSP scenario is unlikely for higgsinos. This would require relatively small but negative $M_1$ values and large $M_2$, see~\cite{Giudice:1995np,Bomark:2013nya}.}.
In Fig.~\ref{fig:gBR} we show the branching ratios of the gluino calculated using {\tt  SUSYHIT 1.4}~\cite{Djouadi:2006bz}, as a function of $\Delta m = m_{\tilde g}-m_{\tilde\chi_1^0}$ for $\tan\beta = 1.5$ (solid lines) and $\tan\beta = 30$ (dashed lines). The gluino mass is fixed at $m_{\tilde g} = 900$ GeV and the other parameters are as given above. 

We see that at least one top quark will be produced per event on average even down to $\Delta m \sim 350$~GeV, where the proximity of  the  $\tilde g \to tb\chi_1^\pm$ threshold becomes important. For $\tan\beta = 30$ we get a sizeable contribution from the decays $\tilde g \to b \bar b \tilde\chi_{1,2}^0$, reducing somewhat the production of top quarks at high $\Delta m$.
We find that lowering $M_1$ and/or $M_2$ down to the gluino mass changes
little: there is a slight increase in the first and second generation quark
decays versus the gluino loop decay below the $tb\chi_1^\pm$ threshold, but no
significant impact above  $\Delta m \sim 350$ GeV. This ensures the presence
of a significant number of events with additional leptons from leptonic top
decays, which we will see have an impact on the allowed parameter space of the model. The decays to first and second generation quarks are heavily suppressed.

The ATLAS analysis assumed equal branching fractions of $\tilde g \rightarrow
qq \tilde\chi_1^0$ for $q=u,d,c,s$, ignoring the heavy quark decays. This
simplifying assumption has relatively little impact on their analysis and the
bounds set because of the focus on leptons from the $Z$ boson. However, the structure of GGM predicts generic sum rules on the sfermion soft masses~\cite{Meade:2008wd},
\begin{eqnarray}
m^2_Q-2m^2_U+m^2_D-m^2_L+m^2_E  & = & 0 \\
2m^2_Q-m^2_U-m^2_D-2m^2_L+m^2_E & = & 0,
\end{eqnarray}
which makes decoupling only the third generation squarks challenging.\footnote{Technically the ATLAS model does not fulfil these sum rules, however, a slight modification of the soft mass parameters would.} One is faced with the choice of either making the phenomenological assumption that the sum rules are broken somehow, increasing the number of free soft sfermion mass parameters in the model, and requiring an explanation of why the third generation masses are significantly heavier, {\it e.g.}\ one could speculate that something along the lines of higgsed gauge mediation~\cite{Craig:2012yd} could work.\footnote{With first and second generation sfermions at $1.5$~TeV, the third generation sfermions must be raised to $\sim 5$~TeV for the gluino branching ratios to light quarks to equal those to third generation quarks.} Or, to reduce the number of top quarks produced, one can at best decouple all squarks except the lightest bottom squarks by an appropriate choice of the soft masses, however, the decay $\tilde g \to tb\chi_1^\pm$ remains. Neither seems very consistent with the simple model of ATLAS\@. As a result we will here include gluino decays to top and bottom quarks in the model.

\section{Scan and simulations \label{sec:scan}}

We perform a grid scan over the range $0$--$1500$~GeV in $\mu$ and $M_3$, using a step size of $15$~GeV in both directions. At each step we calculate the resulting sparticle spectrum using {\tt SOFTSUSY 3.5.1}~\cite{Allanach:2001kg} and the sparticle branching ratios with {\tt  SUSYHIT 1.4}~\cite{Djouadi:2006bz}.  Spectrum and decay information is communicated via the SUSY Les Houches Accord~\cite{Skands:2003cj}, using {\tt PySLHA}~\cite{Buckley:2013jua}. For all parameter points we check that $m_{\tilde g} > m_{\tilde\chi_1^0}$, and that the NLSP is mostly higgsino (more than 0.90). At each point we generate 100\,000 SUSY Monte Carlo events with gluino pair production using {\tt Pythia 8.186}~\cite{Sjostrand:2006za,Sjostrand:2007gs}. The cross sections used are based on {\tt Prospino}~\cite{Beenakker:1996ch}, using the {\tt NLLfast} software including also NLL re-summation of soft gluon emission~\cite{Kulesza:2008jb,Kulesza:2009kq,Beenakker:2009ha,Beenakker:2011fu}. These events are then propagated through our implementations of several collider analyses, detailed below.

%
%
\begin{table}
\begin{tabular}{cc} \hline
observed & 29 \\
background & 10.6$\pm 3.2$ \\
number of sigma & 3.0 \\
$s$ (95$\%$ CL) & 7.1-31.8 \\
\hline
\end{tabular}
\caption{\label{tab:ev} Summary of {\tt ATLAS\_onZ} constraints, showing the
  observed number of events, the number of expected 
  Standard Model events inferred from data, the number of sigma the excess
  corresponds to and the $95\%$ CL constraint upon a putative number of signal
  events $s$. The first three data are taken from Ref.~\cite{Aad:2015wqa},
  whereas we infer the bound on $s$ ourselves (see text).}
\end{table}
The ATLAS analysis with the excess requires two leading opposite sign same
flavor (OSSF) leptons with $p_T>25,10$~GeV and invariant mass $81< m_{ll} < 101$~GeV,
a minimum missing transverse energy of $E_T^{\rm miss}>225$~GeV, at least two
jets, and total transverse energy $H_T> 600$~GeV, where $H_T$ is given as the
scalar sum of the transverse momenta of the two leading leptons and all accepted
jets. Jets are reconstructed with the anti-$k_T$
algorithm~\cite{Cacciari:2008gp} using {\tt FastJet}~\cite{Cacciari:2011ma},
with a jet radius parameter of $R=0.4$, and are required to have $p_T>35$ GeV
and lie within $|\eta|<2.5$. 
In the following we will denote the signal
region with the sum of $ee$ and $\mu\mu$ events as {\tt ATLAS\_onZ}. 
We summarize the ATLAS measurements in Table~\ref{tab:ev}.
In order to calculate a constraint upon the number of non-Standard Model
signal events $s$, we 
profile over a Gaussian background rate, see Section~\ref{sec:stats}, but
otherwise use Poisson statistics. If other combined constraints predict an {\tt
  ATLAS\_onZ} signal rate outside of this range, we shall conclude that they
are incompatible with the signal at the 95$\%$ CL.

The on-$Z$ CMS analysis~\cite{Khachatryan:2015lwa}, 
here called {\tt CMS\_onZ}, requires a leading pair of opposite sign same flavor 
leptons satisfying $p_T>20$~GeV and $81< m_{ll} < 101$~GeV. 
Three signal regions are constructed, covering the ranges $100$--$200$ GeV, $200$--$300$
GeV and $>\!300$ GeV in $E_T^{\rm miss}$, all requiring at least three jets with 
$p_T>40$~GeV and $|\eta|<3.0$. For jet reconstruction the anti-$k_T$ algorithm 
with $R=0.5$ is used. A notable difference with respect to the event selection 
in {\tt ATLAS\_onZ} is that no cut on $H_T$ is applied. 
While there are some details of the original analysis which are difficult to
reproduce outside of the experimental collaborations, {\it e.g.}\ trigger
efficiencies, by simulating models similar to those used for interpretation in
{\tt ATLAS\_onZ} and {\tt CMS\_onZ} we have checked that our implementations
reproduce the observed limits to within theoretical uncertainties, under the
assumptions made. 

In addition to the leptonic-$Z+E_T^{\rm miss}$ analyses from CMS and ATLAS, the scenario used here could be constrained by other searches involving leptons. This includes three and four lepton final states where extra leptons are produced in leptonic top decays, or from two chains with $Z$s. The latter is heavily suppressed by the leptonic branching ratio of the $Z$, down to $\sim7$\% of the number of events with a single leptonically decaying $Z$, thus of the order of two events could be expected for the given luminosity, depending on the exact cuts of such an analysis.

We check the most relevant searches which are the ATLAS stop search with leptons {\tt ATLAS\_stop\_L100}~\cite{Aad:2014qaa}, and the CMS multi-lepton search with three or four leptons, {\tt CMS\_multilepton}~\cite{Chatrchyan:2014aea}. From the analysis in {\tt ATLAS\_stop\_L100} we include the signal region {\tt L100} requiring exactly two opposite-sign leptons with $p_T>25,10$~GeV, at least two jets with $p_T>100,50$~GeV, a `stransverse mass' $m_{T2} > 100$~GeV and an invariant mass $m_{ll}$ for the two leptons outside the range $71$--$111$~GeV. For the GGM model studied here, the cut on $m_{ll}$ means that {\tt ATLAS\_stop\_L100}  is mainly sensitive to events where neither of the two $Z$s decay leptonically. 

The CMS multi-lepton search requires at least three isolated leptons with $p_T>20,10,10$~GeV within $|\eta|<2.4$. Jets are subject to the requirements $p_T>30$~GeV and $|\eta|<2.5$. Accepted events are divided into a large number of signal regions based on the number of opposite-sign same-flavour lepton pairs, $E_T^{\rm miss}$, the presence of a OSSF lepton pair with an invariant mass in the $75$--$105$~GeV range, the scalar sum of jet $p_T$s and the number of tagged $b$-jets.
Due to an overlap between {\tt CMS\_onZ} and the most relevant signal regions in
{\tt CMS\_multilepton}, in the combination we use {\tt CMS\_multilepton} for
parameter regions where the gluino--neutralino mass difference $\Delta m$ is
larger than $500$~GeV and {\tt CMS\_onZ} for mass differences smaller than
this. 
A choice like this is necessary in order to have statistically independent
signal regions when we do not have the information to take into account the correlations. As can be seen from Fig.~\ref{fig:gBR}, the choice ensures
that {\tt CMS\_multilepton} is only applied in the region of parameter space
where additional leptons can be expected due to a sizeable production of top
quarks. 

Given the small leptonic branching ratios of $W$ and $Z$, the GGM scenario
studied can also be constrained from searches for zero-lepton final states. We
include the signal region {\tt 3j} from the ATLAS search for final states with
jets and missing energy~\cite{Aad:2014wea}, here called {\tt
  ATLAS\_jMET\_3j}. Besides a lepton veto, this signal region requires at
least three jets with $p_T>130,60,60$~GeV and $|\eta| < 2.8$, missing energy
$E_T^{\rm miss}>160$~GeV and an `effective mass' $m_{\rm eff}>2200$~GeV. Also,
the missing energy is required to account for at least 30\% of the effective
mass combination of $E_T^{\rm miss}$ and the three leading jet $p_T$s.  

No significant excesses are seen in either of these searches. For all three
searches we have checked, as above, that we can reproduce the relevant limits
on SUSY interpretations presented by the experiments.

\section{Statistics \label{sec:stats}}
In order to combine the results from all the analyses, each independent signal region $i$ is assigned a likelihood $\mathcal{L}_i$ consisting of a Poisson factor for the total event count and a Gaussian for modelling the background uncertainty:
\begin{equation}
  \mathcal{L}_i(s_i,b_i) = \text{Pois}(n_i|s_i+b_i) \times \text{Gauss}(b_{m_i}|b_i,\sigma_{b_i}).
  \label{eq:llhfactor}
\end{equation}
Here $n_i$ is the observed number of events, $s_i$ and $b_i$ are the expected number of signal and background events, and $b_{m_i}$ is the observed background measurement with an expected standard deviation $\sigma_{b_i}$. Inserting the observed values for $n_i$, $b_{m_i}$ and $\sigma_{b_i}$ we are left with a likelihood function for the two parameters $s_i$ and $b_i$. While $s_i$ will be a function of the SUSY parameters $\mu$ and $M_3$, $b_i$ is an unknown nuisance parameter which we eliminate by profiling $\mathcal{L}_i$ over $b_i$. With all the $\mathcal{L}_i$ coming from independent signal regions, the combined likelihood is then simply given by
\begin{equation}
  \mathcal{L}({\mathbf{s}}) = \prod\limits_i\mathcal{L}_i(s_i,\hat{\hat{b}}_i),
  \label{eq:llhcombine}
\end{equation}
where the double hat indicates that we have maximized $\mathcal{L}_i(s_i,b_i)$ over $b_i$ subject to a fixed value of $s_i$.
%
%

For any given parameter point in $\mu$ and $M_3$, the signal expectation values $\mathbf{s}$ are in principle fully determined. In order to set limits in the model parameter space we introduce a common signal strength parameter $\mu_s$ such that the expected signal yield in signal region $i$ is $\mu_s s_i$. Points in the SUSY parameter space for which the upper limit on $\mu_s$ is found to be less than $1$ will be excluded at the confidence level chosen for the test.

For every choice of the SUSY parameters we now have a single-parameter likelihood function $\mathcal{L}(\mu_s) \equiv \mathcal{L}(\mu_s \mathbf{s})$. From the likelihood ratio
\begin{equation}
  \lambda(\mu_s) = \frac{\mathcal{L}(\mu_s)}{\mathcal{L}(\hat{\mu}_s)},
  \label{eq:llhratio}
\end{equation}
we construct a test statistic $q$ given by
\begin{equation}
  q = 
  \begin{cases}
    -2\ln\lambda(\mu_s) &\hat{\mu}_s \leq \mu_s\\
    0 &\hat{\mu}_s > \mu_s,
  \end{cases}
  \label{eq:qstatistic}
\end{equation}
where $\hat{\mu}_s$ is the value of $\mu_s$ that maximizes $\mathcal{L}(\mu_s)$, {\it i.e.}\ the signal strength value preferred by the observed data.
Higher values of $q$ correspond to increasing disagreement between data and the hypothesized value of $\mu_s$, but only in the direction of $\mu_s > \hat{\mu}_s$. For a given $\mu_s$ the observed value $q_{\text{obs}}$ of $q$ is calculated from the data. The $p$-value for this observation is then found from 
\begin{equation}
  p_{\mu_s} = \int\limits^{\infty}_{q_{\text{obs}}} f(q|\mu_s)\,dq,
  \label{eq:mupval}
\end{equation}
where $f(q|\mu_s)$ is the pdf of $q$. To determine $p_{\mu_s}$ we make use of the asymptotic limit in which $f(q|\mu_s)$ is given by a `half chi-square' distribution, {\it i.e.}\ an equally-weighted sum of a delta function at zero and a chi-square distribution for one degree of freedom~\cite{Cowan:2010js}. The 95\% CL$_{s}$~\cite{Read:2002hq} upper limit on $\mu_s$ is the highest value of $\mu_s$ satisfying
\begin{equation}
  \frac{p_{\mu_s}}{1-p_0} \geq 0.05,
  \label{eq:CLs}
\end{equation}
where $p_0$ is the $p$-value for the test statistic
\begin{equation}
  q_0 = 
  \begin{cases}
    -2\ln\lambda(0) &\hat{\mu}_s \geq 0\\
    0 &\hat{\mu}_s < 0,
  \end{cases}
  \label{eq:q0statistic}
\end{equation}
used to test the level of disagreement between data and the background-only hypothesis.

\section{Results}

\begin{figure}[h!]
  \includegraphics[width=8cm]{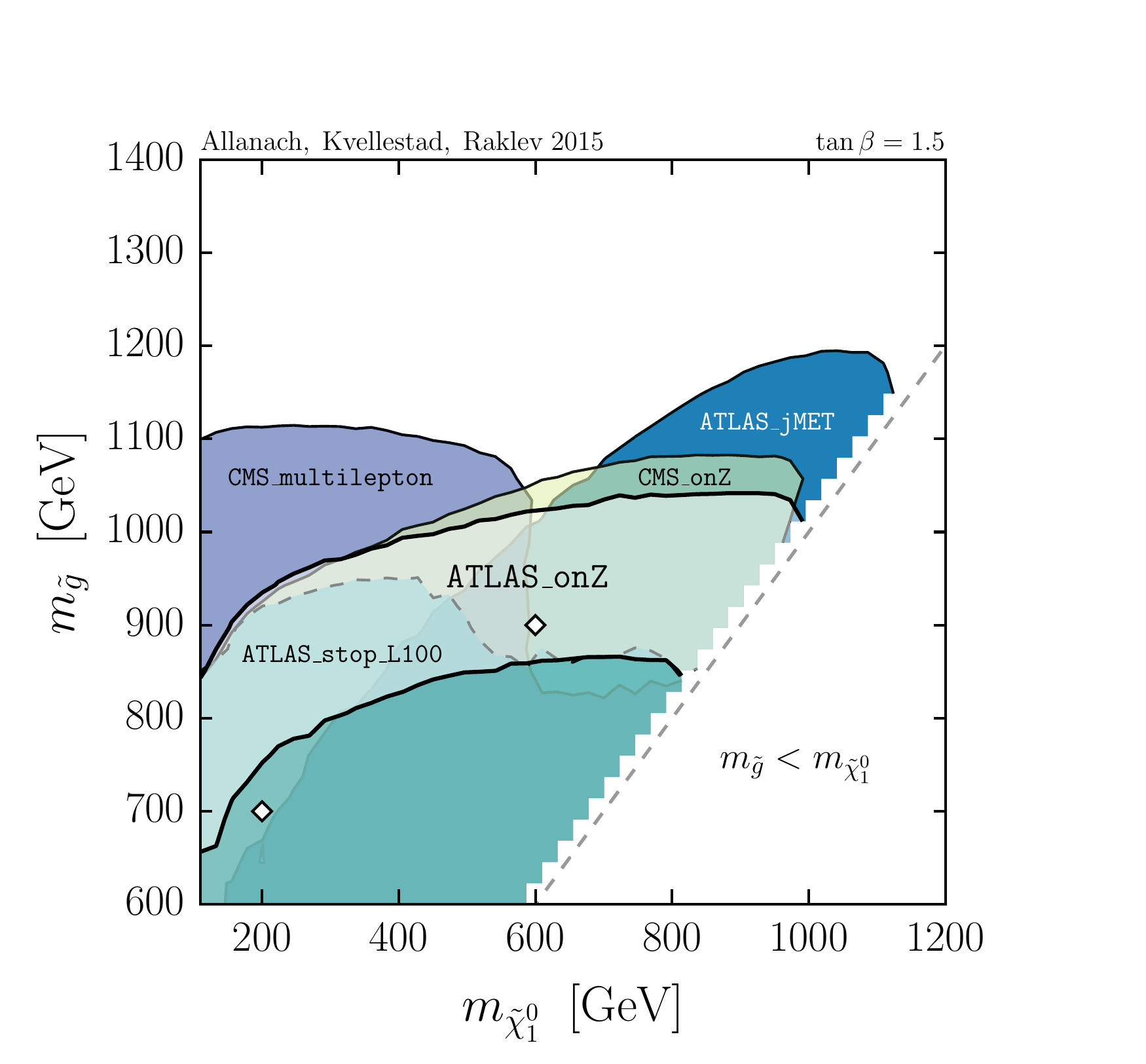}
  \includegraphics[width=8cm]{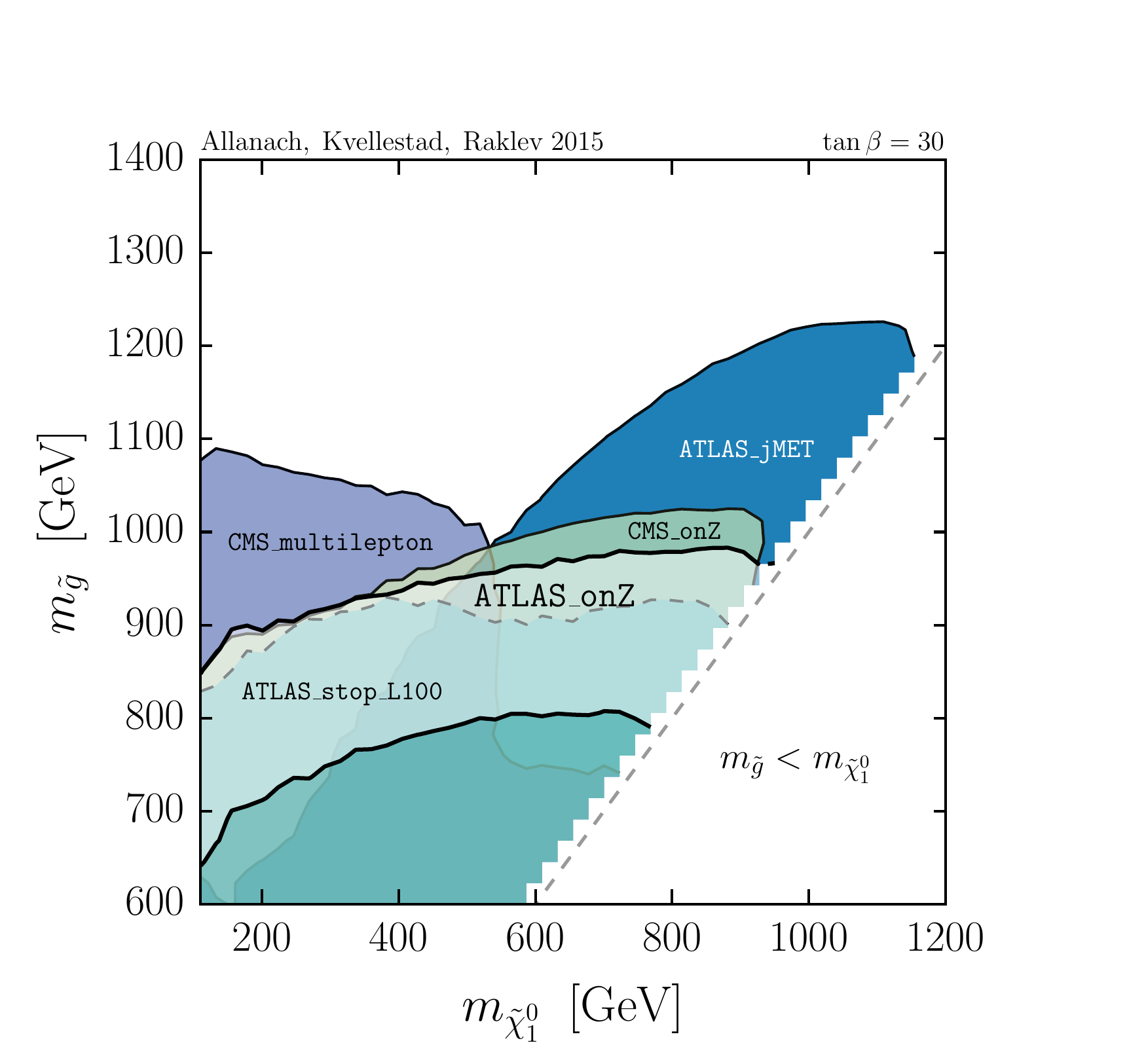}
  \caption{The band in the $(m_{\tilde\chi_1^0},m_{\tilde g})$--plane
    preferred by {\tt ATLAS\_onZ} at 95\% CL (lighter region), compared to the
    coloured 95\% CL$_s$ 
    exclusion contours from {\tt CMS\_onZ}, {\tt ATLAS\_stop\_L100}, {\tt
      CMS\_multilepton} and {\tt ATLAS\_jMET} for $\tan\beta=1.5$ (top),
$\tan \beta=30$ (bottom). The {\tt ATLAS\_stop\_L100} exclusion region
boundary is shown as a dashed line. Two ATLAS benchmark points are indicated with
white diamond markers.}
  \label{fig:all_comb_tanb}
\end{figure}
We show the 95$\%$ CL allowed region for the 3$\sigma$ 
{\tt ATLAS\_onZ} excess in Fig.~\ref{fig:all_comb_tanb} for
$\tan \beta=1.5,30$ as the lighter band. 
The 95$\%$ CL$_s$ excluded regions
from the other searches are overlaid, and for
reference the two ATLAS benchmark points at $(m_{\tilde g},
m_{\tilde\chi_1^0})=(700,200)$~GeV and $(900,600)$~GeV, $\tan\beta=1.5$, are indicated with
white diamond markers. It is clear from the figure that all
points explaining the {\tt ATLAS\_onZ} excess for either value of $\tan \beta$ 
fall afoul of at least one of the other searches. Indeed, {\tt CMS\_onZ} alone
is already incompatible with the excess at the 95$\%$ CL$_s$ level except for
two small regions. One region, around $m_{\tilde \chi_1^0} \approx 950$ GeV,
$m_{\tilde g} \approx 980$ GeV 
which is anyway well excluded by the {\tt ATLAS\_jMET} searches. The other
region, including the point $m_{\tilde \chi_1^0}=190$ GeV, $m_{\tilde g}=930$
GeV is excluded by 
{\tt CMS\_multilepton}.
When $\tan \beta$ is changed, the exclusion contours move somewhat in gluino and
lightest neutralino mass, but the qualitative conclusions remain unchanged. 

\begin{figure}[h!]
  \includegraphics[width=8cm]{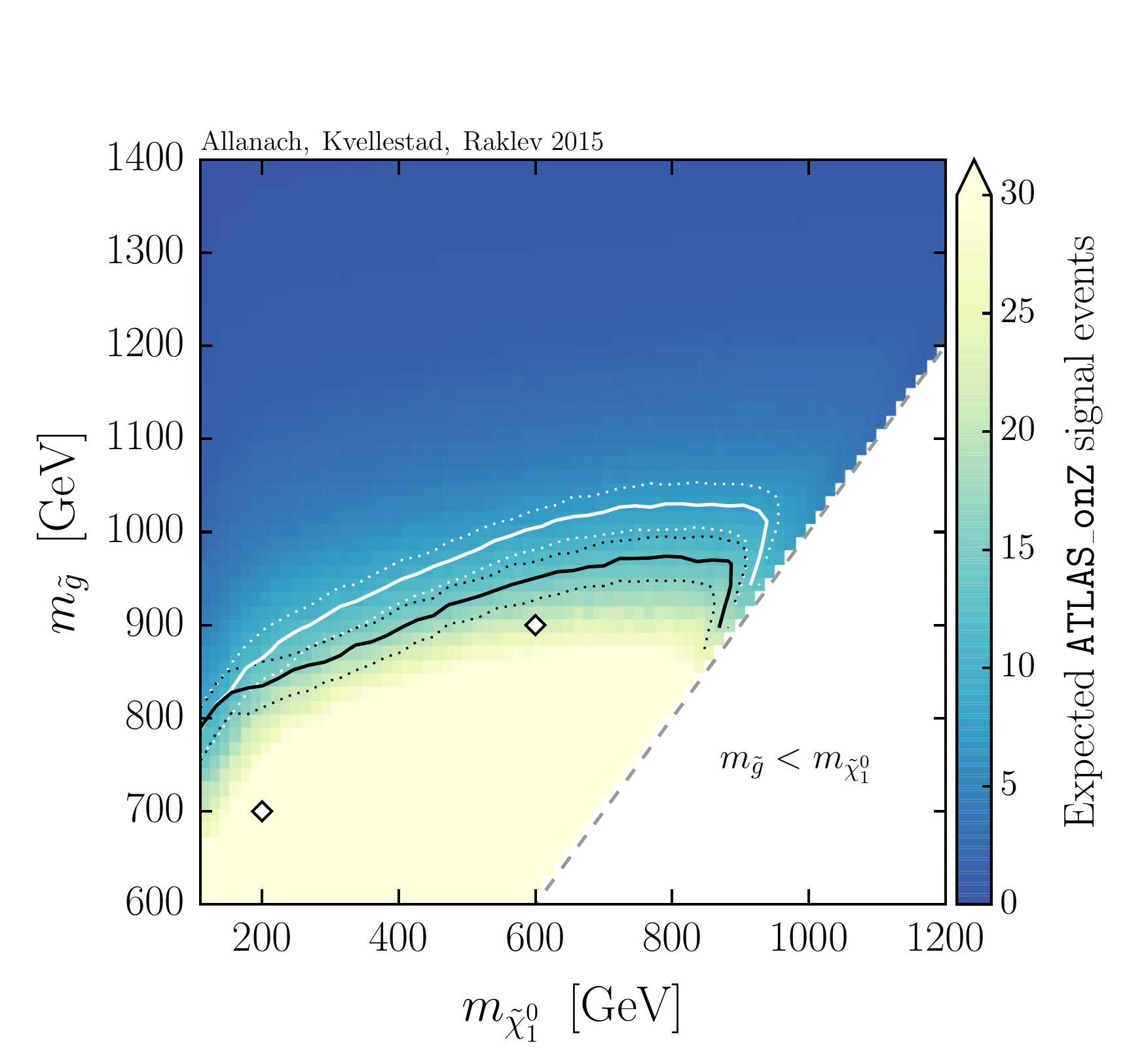}
  \caption{The 95\% CL$_s$ exclusion curves in the
    $(m_{\tilde\chi_1^0},m_{\tilde g})$--plane for $\tan\beta=1.5$ (white) and
    $\tan\beta=30$ (black), using both the {\tt ATLAS\_onZ} and {\tt CMS\_onZ}
    signal regions. The color map shows the expected number of {\tt
      ATLAS\_onZ} signal events for the $\tan\beta=1.5$ scenario. The two
    ATLAS benchmark points are indicated with white diamond markers. The
    region below each curve is excluded by the combination.
    {\tt ATLAS\_onZ} signal events are constrained to be below
    13(12) at the 95$\%$ CL for $\tan \beta=1.5(30)$.
  }
  \label{fig:LLH2}
\end{figure}

Given the tension between the other searches and the {\tt ATLAS\_onZ} excess,
we now combine the {\tt CMS\_onZ} exclusion with various other searches
to see what the combined data set predicts for the number of signal events in
the excess. The 95\% CL$_s$ bound in the $(m_{\tilde\chi_1^0},m_{\tilde g})$--plane
resulting from combining {\tt ATLAS\_onZ} and {\tt CMS\_onZ} is shown in
Fig.~\ref{fig:LLH2}. Here the white contour depicts the limit obtained for
$\tan\beta = 1.5$, while the black contour is for $\tan\beta = 30$. Also shown
are the bounds given a 20\% systematic uncertainty on the cross section. 
The color map shows the predicted number of signal events for the {\tt
  ATLAS\_onZ} analysis in the scenario with $\tan\beta = 1.5$. 
Since the squarks are decoupled from gluino production
here, the production cross section for the model is lower than the original
ATLAS scenario. Still, both ATLAS benchmark points are excluded at the 95\%
confidence level from the combination of {\tt ATLAS\_onZ} and {\tt CMS\_onZ}
alone. 
On the other hand, there are still some points left allowed at the 95$\%$
CL that predict an {\tt ATLAS\_onZ} signal rate of up to 13(12) for $\tan
\beta =1.5(30)$. These are {\em within}\/ the 95$\%$ CL signal rate region of 
$7.1-31.8$ and so are still compatible with the {\tt ATLAS\_onZ} signal at the
95$\%$ CL.

\begin{figure}[h!]
  \includegraphics[width=8cm]{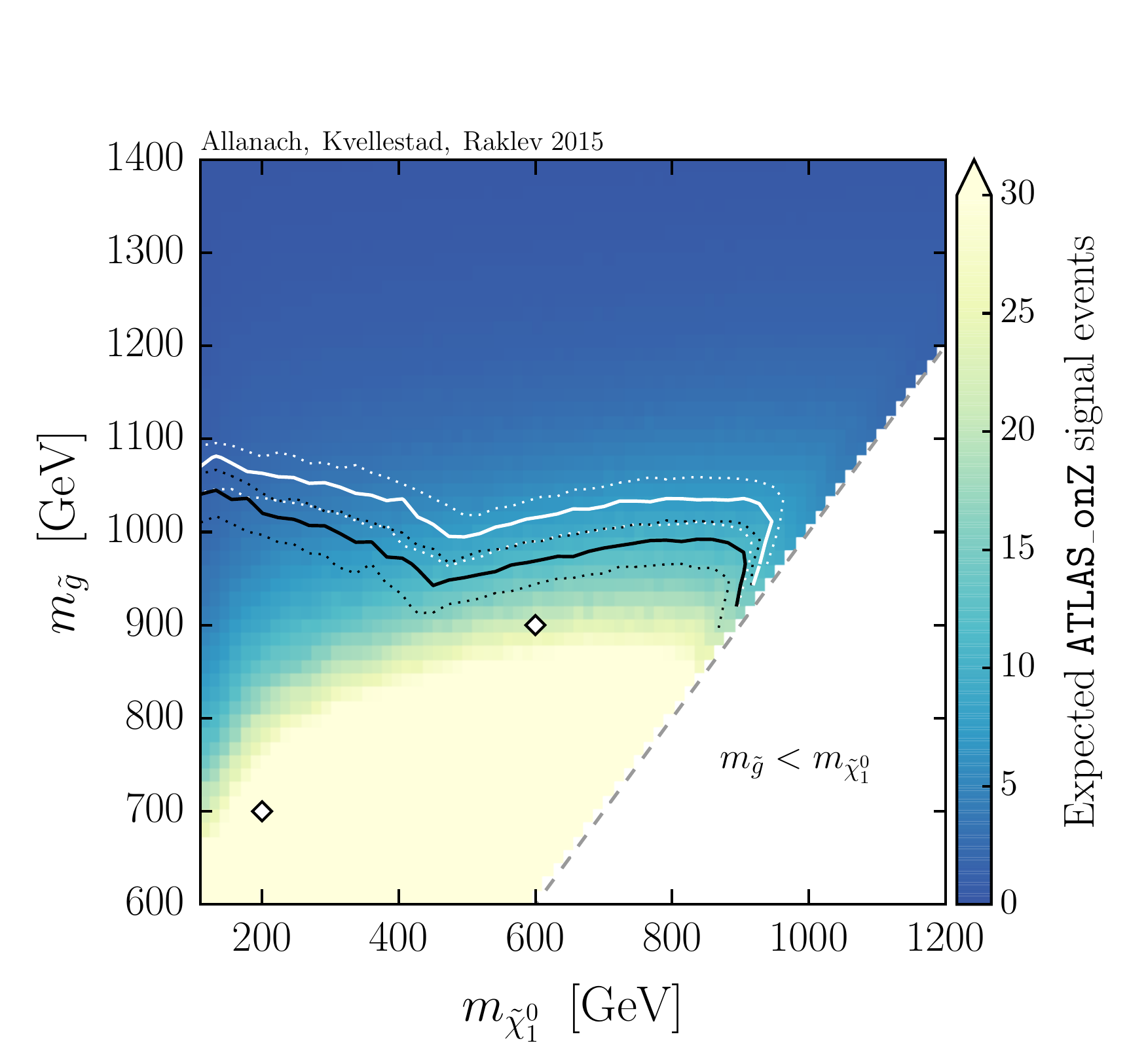}
  \caption{The 95\% CL$_s$ exclusion curves in the
    $(m_{\tilde\chi_1^0},m_{\tilde g})$--plane from combining {\tt
      ATLAS\_onZ}, {\tt CMS\_onZ}, {\tt ATLAS\_stop\_L100} and {\tt
      CMS\_multilepton}.  The color map shows the expected number of {\tt
      ATLAS\_onZ} signal events for the $\tan\beta=1.5$ scenario. The two
    ATLAS benchmark points are indicated with white diamond markers. The
    region below each curve is excluded by the combination. 
    {\tt ATLAS\_onZ} signal events are constrained to be below
    13(11) at the 95$\%$ CL for $\tan \beta=1.5(30)$.
}
  \label{fig:LLH5}
\end{figure}

If we add on the contributions from {\tt ATLAS\_stop\_L100} and {\tt
  CMS\_multilepton} the resulting 95\% CL$_s$ bound is shown in
Fig.~\ref{fig:LLH5}. As expected, the exclusion limits are improved in the
regions of large gluino--neutralino mass difference, where the production of
additional leptons through top quarks is significant. The slight dip in the
contour at $(m_{\tilde g}, m_{\tilde\chi_1^0})\sim(1000,500)$~GeV is where the
domains of the {\tt CMS\_multilepton} and {\tt CMS\_onZ} analyses meet. 
In the region of small gluino--neutralino mass difference the limit remains approximately unchanged,
and the combined allowed region predicts up to 13(11) {\tt ATLAS\_onZ} signal events for $\tan \beta=1.5(30)$,
still consistent with the $7.1-31.8$ 95$\%$ CL constraint.

\begin{figure}[h!]
  \includegraphics[width=8cm]{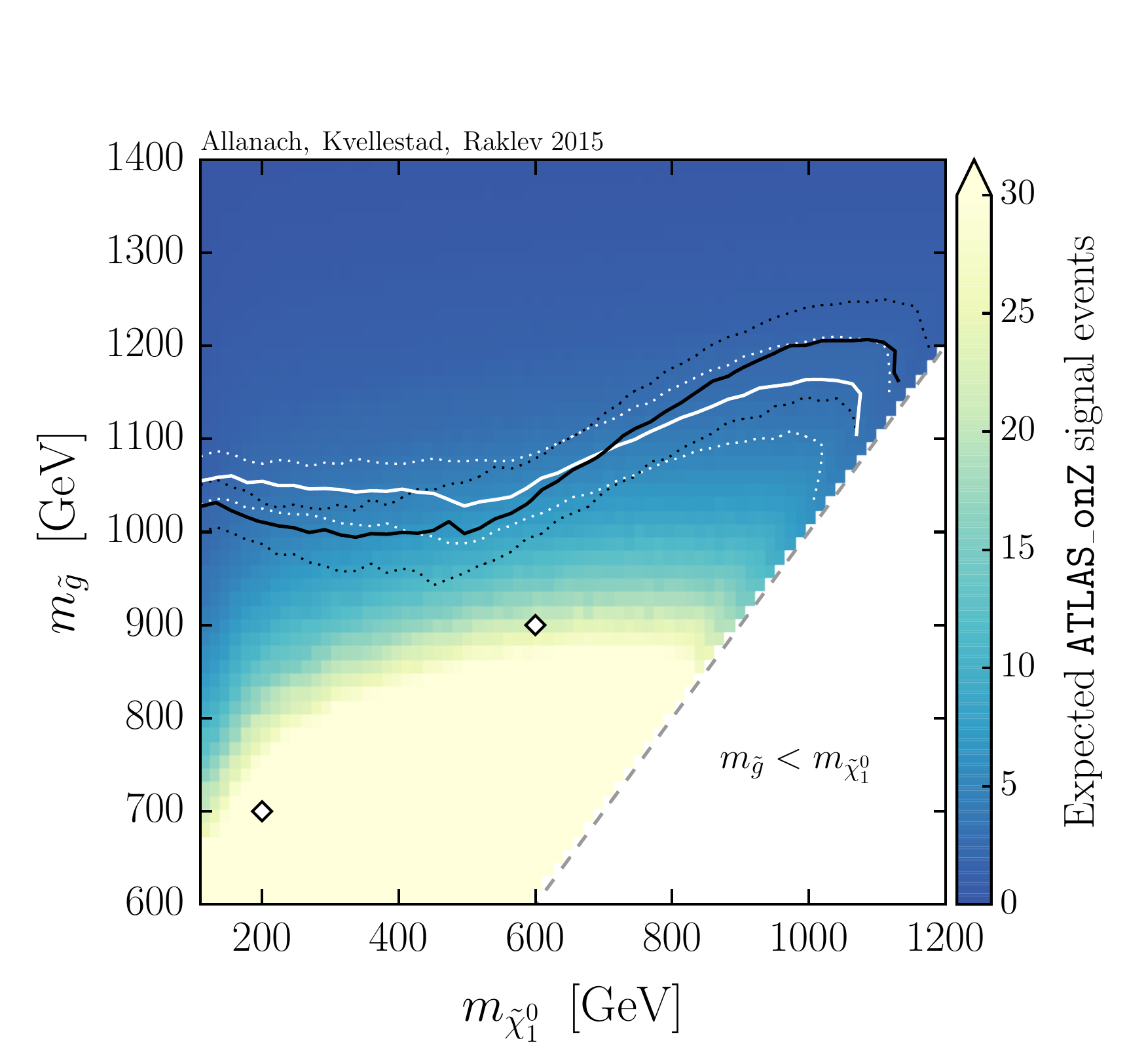}
  \caption{The 95\% CL$_s$ exclusion curves in the
    $(m_{\tilde\chi_1^0},m_{\tilde g})$--plane from combining all collider
    searches detailed in the text.  The color map shows the expected number of {\tt
      ATLAS\_onZ} signal events for the $\tan\beta=1.5$ scenario. The two
    ATLAS benchmark points are indicated with white diamond markers.
The
    region below each curve is excluded by the combination. 
    {\tt ATLAS\_onZ} signal events are constrained to be below
    6(5) at the 95$\%$ CL for $\tan \beta=1.5(30)$.
}
  \label{fig:LLH6}
\end{figure}

The final exclusion limit, obtained after including also {\tt
  ATLAS\_jMET\_3j}, is shown in Fig.~\ref{fig:LLH6}. Due to the lepton veto in
this analysis, the exclusion limit is mainly strengthened in the region with
gluino--neutralino mass differences less than $400$~GeV, where the main source
of leptons is through the small leptonic branching ratio of the $Z$s. With
$\tan\beta = 30$ this effect is further enhanced by the reduced branching
ratio into $Z$s. We note that gluino masses below $1$~TeV are fully excluded
for both values of $\tan\beta$, and the remaining allowed parameter space has
a maximum of $6$ expected signal events for {\tt ATLAS\_onZ}, which is far
from explaining the observed excess. 

The observed excess in {\tt ATLAS\_onZ} was also recently interpreted within a
GGM framework in~\cite{Barenboim:2015afa}, where two new benchmark points are
presented. The main difference with respect to the ATLAS scenario is heavier
squarks, and that the lightest neutralino is a wino--bino mixture. The first
point, referred to as {\tt GGM1}, has $m_{\tilde{g}} = 1088$~GeV,
$m_{\tilde\chi_1^0} = 428$~GeV and squark masses around $2800$~GeV. 
The second point, {\tt GGM2}, has a higher production cross section due to lighter gluinos and squarks, at $911$~GeV and $\sim 2400$~GeV, respectively.
Combining the collider constraints considered here we find that {\tt GGM2} is
excluded at the 95\% confidence level, while {\tt GGM1} escapes
exclusion. However, it should be noted that {\tt GGM1} predicts only $\sim 3$
signal events for {\tt ATLAS\_onZ}\footnote{We note that this is in slight
  disagreement with the simulation in~\cite{Barenboim:2015afa}, which finds $6
  \pm 1$ predicted signal events for {\tt GGM1}. However, this discrepancy
  does not change the conclusion.}. 

One may ask: can one tweak the simplified model in order to squeeze around the
constraints? In our analysis, we have set the simplified model up in order to
{\em 
  maximize}\/ the {\tt ATLAS\_onZ} region compared to the other constraining
searches. If $\tan \beta$ is increased further, the {\tt ATLAS\_onZ} signal
decreases for the same gluino/lightest neutralino masses. Thus, these would
have to be lowered in order to get a signal to fit, and such lighter
sparticles would suffer more from the other searches. If we were to make
squarks lighter, although the {\tt ATLAS\_onZ} signal would increase for the
same gluino/lightest neutralino masses, the {\tt ATLAS\_jMET} constraints
would become {\em much}\/ stronger. In any case, {\tt CMS\_onZ} is in tension
with nearly all of the {\tt ATLAS\_onZ} parameter space, and this is unlikely
to change. 
The {\tt CMS\_multilepton} constraint is mainly due to events with one
leptonically decaying $Z$-boson plus additional leptons from decaying top quarks, 
predicted by the signal model due to the higgsino nature of the neutralino.
If the model is manipulated to reduce the production of such additional leptons,
the weakening of the {\tt CMS\_multilepton} constraint will be compensated by a corresponding
strengthening of {\tt ATLAS\_jMET} as more events will pass the lepton veto.
Thus, although we have not exhaustively covered the full
multi-dimensional MSSM plus light gravitino space, there are good grounds for
expecting that our conclusions --- that the  {\tt ATLAS\_onZ} excess is incompatible with
other searches at the 95$\%$ CL for GGM-type models --- also apply to the full MSSM space. 

\section{Conclusions}
In conclusion, we have seen that a simplified GGM model with only $M_3$ and
$\mu$ as free parameters, and with $\tilde g$, higgsino  $\tilde\chi_1^0$,
$\tilde\chi_2^0$ and $\tilde\chi_1^\pm$, and a gravitino LSP as the only
sparticles produced at the LHC, cannot explain the ATLAS excess reported
in~\cite{Aad:2015wqa} when faced with results from other current new physics
searches. 
Strong bounds on the model that can be set from other leptonic
searches are due to the higgsino nature of the NLSP, leading to the production
of top quarks with leptonic decays. Zero-lepton searches also provide strong
constraints in the parameter regions where leptons are mainly produced through
the small leptonic branching ratio of the $Z$. 
Tension between {\tt ATLAS\_onZ} and the other relevant searches is
evident in Fig.~\ref{fig:all_comb_tanb}. A combined fit to all 
constraints, {\em including the excess}, at the 95$\%$ CL predicts less than 6
signal events in the {\tt   ATLAS\_onZ} search region, compared to 7.1-31.8
being inferred from the {\tt ATLAS\_onZ} search region alone.

\acknowledgements
This work has been partially supported by STFC grant ST/L000385/1. We
thank the Cambridge SUSY Working Group for stimulating discussions. The CPU
intensive parts of this work was performed on the Abel Cluster, owned by the
University of Oslo and the Norwegian meta-center for High Performance Computing
(NOTUR), and operated by the Research Computing Services group at USIT, the
University of Oslo IT-department. The computing time was given by NOTUR
allocation NN9284K, financed through the Research Council of Norway. 

\bibliography{ATLASZ}

\begin{thebibliography}{33}%
\makeatletter
\providecommand \@ifxundefined [1]{%
 \@ifx{#1\undefined}
}%
\providecommand \@ifnum [1]{%
 \ifnum #1\expandafter \@firstoftwo
 \else \expandafter \@secondoftwo
 \fi
}%
\providecommand \@ifx [1]{%
 \ifx #1\expandafter \@firstoftwo
 \else \expandafter \@secondoftwo
 \fi
}%
\providecommand \natexlab [1]{#1}%
\providecommand \enquote  [1]{``#1''}%
\providecommand \bibnamefont  [1]{#1}%
\providecommand \bibfnamefont [1]{#1}%
\providecommand \citenamefont [1]{#1}%
\providecommand \href@noop [0]{\@secondoftwo}%
\providecommand \href [0]{\begingroup \@sanitize@url \@href}%
\providecommand \@href[1]{\@@startlink{#1}\@@href}%
\providecommand \@@href[1]{\endgroup#1\@@endlink}%
\providecommand \@sanitize@url [0]{\catcode `\\12\catcode `\$12\catcode
  `\&12\catcode `\#12\catcode `\^12\catcode `\_12\catcode `\%12\relax}%
\providecommand \@@startlink[1]{}%
\providecommand \@@endlink[0]{}%
\providecommand \url  [0]{\begingroup\@sanitize@url \@url }%
\providecommand \@url [1]{\endgroup\@href {#1}{\urlprefix }}%
\providecommand \urlprefix  [0]{URL }%
\providecommand \Eprint [0]{\href }%
\providecommand \doibase [0]{http://dx.doi.org/}%
\providecommand \selectlanguage [0]{\@gobble}%
\providecommand \bibinfo  [0]{\@secondoftwo}%
\providecommand \bibfield  [0]{\@secondoftwo}%
\providecommand \translation [1]{[#1]}%
\providecommand \BibitemOpen [0]{}%
\providecommand \bibitemStop [0]{}%
\providecommand \bibitemNoStop [0]{.\EOS\space}%
\providecommand \EOS [0]{\spacefactor3000\relax}%
\providecommand \BibitemShut  [1]{\csname bibitem#1\endcsname}%
\let\auto@bib@innerbib\@empty
\bibitem [{\citenamefont {Aad}\ \emph {et~al.}(2015{\natexlab{a}})\citenamefont
  {Aad} \emph {et~al.}}]{Aad:2015wqa}%
  \BibitemOpen
  \bibfield  {author} {\bibinfo {author} {\bibfnamefont {G.}~\bibnamefont
  {Aad}} \emph {et~al.} (\bibinfo {collaboration} {ATLAS}),\ }\href@noop {} {\
  (\bibinfo {year} {2015}{\natexlab{a}})},\ \Eprint
  {http://arxiv.org/abs/1503.03290} {arXiv:1503.03290 [hep-ex]} \BibitemShut
  {NoStop}%
\bibitem [{\citenamefont {Khachatryan}\ \emph {et~al.}(2015)\citenamefont
  {Khachatryan} \emph {et~al.}}]{Khachatryan:2015lwa}%
  \BibitemOpen
  \bibfield  {author} {\bibinfo {author} {\bibfnamefont {V.}~\bibnamefont
  {Khachatryan}} \emph {et~al.} (\bibinfo {collaboration} {CMS}),\ }\href@noop
  {} {\  (\bibinfo {year} {2015})},\ \Eprint {http://arxiv.org/abs/1502.06031}
  {arXiv:1502.06031 [hep-ex]} \BibitemShut {NoStop}%
\bibitem [{\citenamefont {Meade}\ \emph {et~al.}(2009)\citenamefont {Meade},
  \citenamefont {Seiberg},\ and\ \citenamefont {Shih}}]{Meade:2008wd}%
  \BibitemOpen
  \bibfield  {author} {\bibinfo {author} {\bibfnamefont {P.}~\bibnamefont
  {Meade}}, \bibinfo {author} {\bibfnamefont {N.}~\bibnamefont {Seiberg}}, \
  and\ \bibinfo {author} {\bibfnamefont {D.}~\bibnamefont {Shih}},\ }\href
  {\doibase 10.1143/PTPS.177.143} {\bibfield  {journal} {\bibinfo  {journal}
  {Prog.Theor.Phys.Suppl.}\ }\textbf {\bibinfo {volume} {177}},\ \bibinfo
  {pages} {143} (\bibinfo {year} {2009})},\ \Eprint
  {http://arxiv.org/abs/0801.3278} {arXiv:0801.3278 [hep-ph]} \BibitemShut
  {NoStop}%
\bibitem [{\citenamefont {Ruderman}\ and\ \citenamefont
  {Shih}(2012)}]{Ruderman:2011vv}%
  \BibitemOpen
  \bibfield  {author} {\bibinfo {author} {\bibfnamefont {J.~T.}\ \bibnamefont
  {Ruderman}}\ and\ \bibinfo {author} {\bibfnamefont {D.}~\bibnamefont
  {Shih}},\ }\href {\doibase 10.1007/JHEP08(2012)159} {\bibfield  {journal}
  {\bibinfo  {journal} {JHEP}\ }\textbf {\bibinfo {volume} {1208}},\ \bibinfo
  {pages} {159} (\bibinfo {year} {2012})},\ \Eprint
  {http://arxiv.org/abs/1103.6083} {arXiv:1103.6083 [hep-ph]} \BibitemShut
  {NoStop}%
\bibitem [{\citenamefont {Ellwanger}(2015)}]{Ellwanger:2015hva}%
  \BibitemOpen
  \bibfield  {author} {\bibinfo {author} {\bibfnamefont {U.}~\bibnamefont
  {Ellwanger}},\ }\href@noop {} {\  (\bibinfo {year} {2015})},\ \Eprint
  {http://arxiv.org/abs/1504.02244} {arXiv:1504.02244 [hep-ph]} \BibitemShut
  {NoStop}%
\bibitem [{\citenamefont {Barenboim}\ \emph {et~al.}(2015)\citenamefont
  {Barenboim}, \citenamefont {Bernabeu}, \citenamefont {Mitsou}, \citenamefont
  {Romero}, \citenamefont {Torro} \emph {et~al.}}]{Barenboim:2015afa}%
  \BibitemOpen
  \bibfield  {author} {\bibinfo {author} {\bibfnamefont {G.}~\bibnamefont
  {Barenboim}}, \bibinfo {author} {\bibfnamefont {J.}~\bibnamefont {Bernabeu}},
  \bibinfo {author} {\bibfnamefont {V.}~\bibnamefont {Mitsou}}, \bibinfo
  {author} {\bibfnamefont {E.}~\bibnamefont {Romero}}, \bibinfo {author}
  {\bibfnamefont {E.}~\bibnamefont {Torro}},  \emph {et~al.},\ }\href@noop {}
  {\  (\bibinfo {year} {2015})},\ \Eprint {http://arxiv.org/abs/1503.04184}
  {arXiv:1503.04184 [hep-ph]} \BibitemShut {NoStop}%
\bibitem [{\citenamefont {Vignaroli}(2015)}]{Vignaroli:2015ama}%
  \BibitemOpen
  \bibfield  {author} {\bibinfo {author} {\bibfnamefont {N.}~\bibnamefont
  {Vignaroli}},\ }\href@noop {} {\  (\bibinfo {year} {2015})},\ \Eprint
  {http://arxiv.org/abs/1504.01768} {arXiv:1504.01768 [hep-ph]} \BibitemShut
  {NoStop}%
\bibitem [{\citenamefont {Meade}\ \emph {et~al.}(2010)\citenamefont {Meade},
  \citenamefont {Reece},\ and\ \citenamefont {Shih}}]{Meade:2009qv}%
  \BibitemOpen
  \bibfield  {author} {\bibinfo {author} {\bibfnamefont {P.}~\bibnamefont
  {Meade}}, \bibinfo {author} {\bibfnamefont {M.}~\bibnamefont {Reece}}, \ and\
  \bibinfo {author} {\bibfnamefont {D.}~\bibnamefont {Shih}},\ }\href {\doibase
  10.1007/JHEP05(2010)105} {\bibfield  {journal} {\bibinfo  {journal} {JHEP}\
  }\textbf {\bibinfo {volume} {1005}},\ \bibinfo {pages} {105} (\bibinfo {year}
  {2010})},\ \Eprint {http://arxiv.org/abs/0911.4130} {arXiv:0911.4130
  [hep-ph]} \BibitemShut {NoStop}%
\bibitem [{\citenamefont {Ambrosanio}\ \emph {et~al.}(1996)\citenamefont
  {Ambrosanio}, \citenamefont {Kane}, \citenamefont {Kribs}, \citenamefont
  {Martin},\ and\ \citenamefont {Mrenna}}]{Ambrosanio:1996jn}%
  \BibitemOpen
  \bibfield  {author} {\bibinfo {author} {\bibfnamefont {S.}~\bibnamefont
  {Ambrosanio}}, \bibinfo {author} {\bibfnamefont {G.~L.}\ \bibnamefont
  {Kane}}, \bibinfo {author} {\bibfnamefont {G.~D.}\ \bibnamefont {Kribs}},
  \bibinfo {author} {\bibfnamefont {S.~P.}\ \bibnamefont {Martin}}, \ and\
  \bibinfo {author} {\bibfnamefont {S.}~\bibnamefont {Mrenna}},\ }\href
  {\doibase 10.1103/PhysRevD.54.5395} {\bibfield  {journal} {\bibinfo
  {journal} {Phys.Rev.}\ }\textbf {\bibinfo {volume} {D54}},\ \bibinfo {pages}
  {5395} (\bibinfo {year} {1996})},\ \Eprint
  {http://arxiv.org/abs/hep-ph/9605398} {arXiv:hep-ph/9605398 [hep-ph]}
  \BibitemShut {NoStop}%
\bibitem [{\citenamefont {Dimopoulos}\ \emph {et~al.}(1997)\citenamefont
  {Dimopoulos}, \citenamefont {Thomas},\ and\ \citenamefont
  {Wells}}]{Dimopoulos:1996yq}%
  \BibitemOpen
  \bibfield  {author} {\bibinfo {author} {\bibfnamefont {S.}~\bibnamefont
  {Dimopoulos}}, \bibinfo {author} {\bibfnamefont {S.~D.}\ \bibnamefont
  {Thomas}}, \ and\ \bibinfo {author} {\bibfnamefont {J.~D.}\ \bibnamefont
  {Wells}},\ }\href {\doibase 10.1016/S0550-3213(97)00030-8} {\bibfield
  {journal} {\bibinfo  {journal} {Nucl.Phys.}\ }\textbf {\bibinfo {volume}
  {B488}},\ \bibinfo {pages} {39} (\bibinfo {year} {1997})},\ \Eprint
  {http://arxiv.org/abs/hep-ph/9609434} {arXiv:hep-ph/9609434 [hep-ph]}
  \BibitemShut {NoStop}%
\bibitem [{\citenamefont {Aad}\ \emph {et~al.}(2015{\natexlab{b}})\citenamefont
  {Aad} \emph {et~al.}}]{Aad:2015zhl}%
  \BibitemOpen
  \bibfield  {author} {\bibinfo {author} {\bibfnamefont {G.}~\bibnamefont
  {Aad}} \emph {et~al.} (\bibinfo {collaboration} {CMS s}),\ }\href@noop {} {\
  (\bibinfo {year} {2015}{\natexlab{b}})},\ \Eprint
  {http://arxiv.org/abs/1503.07589} {arXiv:1503.07589 [hep-ex]} \BibitemShut
  {NoStop}%
\bibitem [{\citenamefont {Dine}\ \emph {et~al.}(2007)\citenamefont {Dine},
  \citenamefont {Seiberg},\ and\ \citenamefont {Thomas}}]{Dine:2007xi}%
  \BibitemOpen
  \bibfield  {author} {\bibinfo {author} {\bibfnamefont {M.}~\bibnamefont
  {Dine}}, \bibinfo {author} {\bibfnamefont {N.}~\bibnamefont {Seiberg}}, \
  and\ \bibinfo {author} {\bibfnamefont {S.}~\bibnamefont {Thomas}},\ }\href
  {\doibase 10.1103/PhysRevD.76.095004} {\bibfield  {journal} {\bibinfo
  {journal} {Phys.Rev.}\ }\textbf {\bibinfo {volume} {D76}},\ \bibinfo {pages}
  {095004} (\bibinfo {year} {2007})},\ \Eprint {http://arxiv.org/abs/0707.0005}
  {arXiv:0707.0005 [hep-ph]} \BibitemShut {NoStop}%
\bibitem [{\citenamefont {Giudice}\ and\ \citenamefont
  {Pomarol}(1996)}]{Giudice:1995np}%
  \BibitemOpen
  \bibfield  {author} {\bibinfo {author} {\bibfnamefont {G.~F.}\ \bibnamefont
  {Giudice}}\ and\ \bibinfo {author} {\bibfnamefont {A.}~\bibnamefont
  {Pomarol}},\ }\href {\doibase 10.1016/0370-2693(96)00060-3} {\bibfield
  {journal} {\bibinfo  {journal} {Phys.Lett.}\ }\textbf {\bibinfo {volume}
  {B372}},\ \bibinfo {pages} {253} (\bibinfo {year} {1996})},\ \Eprint
  {http://arxiv.org/abs/hep-ph/9512337} {arXiv:hep-ph/9512337 [hep-ph]}
  \BibitemShut {NoStop}%
\bibitem [{\citenamefont {Bomark}\ \emph {et~al.}(2014)\citenamefont {Bomark},
  \citenamefont {Kvellestad}, \citenamefont {Lola}, \citenamefont {Osland},\
  and\ \citenamefont {Raklev}}]{Bomark:2013nya}%
  \BibitemOpen
  \bibfield  {author} {\bibinfo {author} {\bibfnamefont {N.-E.}\ \bibnamefont
  {Bomark}}, \bibinfo {author} {\bibfnamefont {A.}~\bibnamefont {Kvellestad}},
  \bibinfo {author} {\bibfnamefont {S.}~\bibnamefont {Lola}}, \bibinfo {author}
  {\bibfnamefont {P.}~\bibnamefont {Osland}}, \ and\ \bibinfo {author}
  {\bibfnamefont {A.}~\bibnamefont {Raklev}},\ }\href {\doibase
  10.1007/JHEP05(2014)007} {\bibfield  {journal} {\bibinfo  {journal} {JHEP}\
  }\textbf {\bibinfo {volume} {1405}},\ \bibinfo {pages} {007} (\bibinfo {year}
  {2014})},\ \Eprint {http://arxiv.org/abs/1310.2788} {arXiv:1310.2788
  [hep-ph]} \BibitemShut {NoStop}%
\bibitem [{\citenamefont {Djouadi}\ \emph {et~al.}(2007)\citenamefont
  {Djouadi}, \citenamefont {Muhlleitner},\ and\ \citenamefont
  {Spira}}]{Djouadi:2006bz}%
  \BibitemOpen
  \bibfield  {author} {\bibinfo {author} {\bibfnamefont {A.}~\bibnamefont
  {Djouadi}}, \bibinfo {author} {\bibfnamefont {M.}~\bibnamefont
  {Muhlleitner}}, \ and\ \bibinfo {author} {\bibfnamefont {M.}~\bibnamefont
  {Spira}},\ }\href@noop {} {\bibfield  {journal} {\bibinfo  {journal} {Acta
  Phys.Polon.}\ }\textbf {\bibinfo {volume} {B38}},\ \bibinfo {pages} {635}
  (\bibinfo {year} {2007})},\ \Eprint {http://arxiv.org/abs/hep-ph/0609292}
  {arXiv:hep-ph/0609292 [hep-ph]} \BibitemShut {NoStop}%
\bibitem [{\citenamefont {Craig}\ \emph {et~al.}(2012)\citenamefont {Craig},
  \citenamefont {McCullough},\ and\ \citenamefont {Thaler}}]{Craig:2012yd}%
  \BibitemOpen
  \bibfield  {author} {\bibinfo {author} {\bibfnamefont {N.}~\bibnamefont
  {Craig}}, \bibinfo {author} {\bibfnamefont {M.}~\bibnamefont {McCullough}}, \
  and\ \bibinfo {author} {\bibfnamefont {J.}~\bibnamefont {Thaler}},\ }\href
  {\doibase 10.1007/JHEP03(2012)049} {\bibfield  {journal} {\bibinfo  {journal}
  {JHEP}\ }\textbf {\bibinfo {volume} {1203}},\ \bibinfo {pages} {049}
  (\bibinfo {year} {2012})},\ \Eprint {http://arxiv.org/abs/1201.2179}
  {arXiv:1201.2179 [hep-ph]} \BibitemShut {NoStop}%
\bibitem [{\citenamefont {Allanach}(2002)}]{Allanach:2001kg}%
  \BibitemOpen
  \bibfield  {author} {\bibinfo {author} {\bibfnamefont {B.}~\bibnamefont
  {Allanach}},\ }\href {\doibase 10.1016/S0010-4655(01)00460-X} {\bibfield
  {journal} {\bibinfo  {journal} {Comput.Phys.Commun.}\ }\textbf {\bibinfo
  {volume} {143}},\ \bibinfo {pages} {305} (\bibinfo {year} {2002})},\ \Eprint
  {http://arxiv.org/abs/hep-ph/0104145} {arXiv:hep-ph/0104145 [hep-ph]}
  \BibitemShut {NoStop}%
\bibitem [{\citenamefont {Skands}\ \emph {et~al.}(2004)\citenamefont {Skands},
  \citenamefont {Allanach}, \citenamefont {Baer}, \citenamefont {Balazs},
  \citenamefont {Belanger} \emph {et~al.}}]{Skands:2003cj}%
  \BibitemOpen
  \bibfield  {author} {\bibinfo {author} {\bibfnamefont {P.~Z.}\ \bibnamefont
  {Skands}}, \bibinfo {author} {\bibfnamefont {B.}~\bibnamefont {Allanach}},
  \bibinfo {author} {\bibfnamefont {H.}~\bibnamefont {Baer}}, \bibinfo {author}
  {\bibfnamefont {C.}~\bibnamefont {Balazs}}, \bibinfo {author} {\bibfnamefont
  {G.}~\bibnamefont {Belanger}},  \emph {et~al.},\ }\href {\doibase
  10.1088/1126-6708/2004/07/036} {\bibfield  {journal} {\bibinfo  {journal}
  {JHEP}\ }\textbf {\bibinfo {volume} {0407}},\ \bibinfo {pages} {036}
  (\bibinfo {year} {2004})},\ \Eprint {http://arxiv.org/abs/hep-ph/0311123}
  {arXiv:hep-ph/0311123 [hep-ph]} \BibitemShut {NoStop}%
\bibitem [{\citenamefont {Buckley}(2013)}]{Buckley:2013jua}%
  \BibitemOpen
  \bibfield  {author} {\bibinfo {author} {\bibfnamefont {A.}~\bibnamefont
  {Buckley}},\ }\href@noop {} {\  (\bibinfo {year} {2013})},\ \Eprint
  {http://arxiv.org/abs/1305.4194} {arXiv:1305.4194 [hep-ph]} \BibitemShut
  {NoStop}%
\bibitem [{\citenamefont {Sjostrand}\ \emph {et~al.}(2006)\citenamefont
  {Sjostrand}, \citenamefont {Mrenna},\ and\ \citenamefont
  {Skands}}]{Sjostrand:2006za}%
  \BibitemOpen
  \bibfield  {author} {\bibinfo {author} {\bibfnamefont {T.}~\bibnamefont
  {Sjostrand}}, \bibinfo {author} {\bibfnamefont {S.}~\bibnamefont {Mrenna}}, \
  and\ \bibinfo {author} {\bibfnamefont {P.~Z.}\ \bibnamefont {Skands}},\
  }\href {\doibase 10.1088/1126-6708/2006/05/026} {\bibfield  {journal}
  {\bibinfo  {journal} {JHEP}\ }\textbf {\bibinfo {volume} {0605}},\ \bibinfo
  {pages} {026} (\bibinfo {year} {2006})},\ \Eprint
  {http://arxiv.org/abs/hep-ph/0603175} {arXiv:hep-ph/0603175 [hep-ph]}
  \BibitemShut {NoStop}%
\bibitem [{\citenamefont {Sjostrand}\ \emph {et~al.}(2008)\citenamefont
  {Sjostrand}, \citenamefont {Mrenna},\ and\ \citenamefont
  {Skands}}]{Sjostrand:2007gs}%
  \BibitemOpen
  \bibfield  {author} {\bibinfo {author} {\bibfnamefont {T.}~\bibnamefont
  {Sjostrand}}, \bibinfo {author} {\bibfnamefont {S.}~\bibnamefont {Mrenna}}, \
  and\ \bibinfo {author} {\bibfnamefont {P.~Z.}\ \bibnamefont {Skands}},\
  }\href {\doibase 10.1016/j.cpc.2008.01.036} {\bibfield  {journal} {\bibinfo
  {journal} {Comput.Phys.Commun.}\ }\textbf {\bibinfo {volume} {178}},\
  \bibinfo {pages} {852} (\bibinfo {year} {2008})},\ \Eprint
  {http://arxiv.org/abs/0710.3820} {arXiv:0710.3820 [hep-ph]} \BibitemShut
  {NoStop}%
\bibitem [{\citenamefont {Beenakker}\ \emph {et~al.}(1997)\citenamefont
  {Beenakker}, \citenamefont {Hopker}, \citenamefont {Spira},\ and\
  \citenamefont {Zerwas}}]{Beenakker:1996ch}%
  \BibitemOpen
  \bibfield  {author} {\bibinfo {author} {\bibfnamefont {W.}~\bibnamefont
  {Beenakker}}, \bibinfo {author} {\bibfnamefont {R.}~\bibnamefont {Hopker}},
  \bibinfo {author} {\bibfnamefont {M.}~\bibnamefont {Spira}}, \ and\ \bibinfo
  {author} {\bibfnamefont {P.}~\bibnamefont {Zerwas}},\ }\href {\doibase
  10.1016/S0550-3213(97)80027-2} {\bibfield  {journal} {\bibinfo  {journal}
  {Nucl.Phys.}\ }\textbf {\bibinfo {volume} {B492}},\ \bibinfo {pages} {51}
  (\bibinfo {year} {1997})},\ \Eprint {http://arxiv.org/abs/hep-ph/9610490}
  {arXiv:hep-ph/9610490 [hep-ph]} \BibitemShut {NoStop}%
\bibitem [{\citenamefont {Kulesza}\ and\ \citenamefont
  {Motyka}(2009{\natexlab{a}})}]{Kulesza:2008jb}%
  \BibitemOpen
  \bibfield  {author} {\bibinfo {author} {\bibfnamefont {A.}~\bibnamefont
  {Kulesza}}\ and\ \bibinfo {author} {\bibfnamefont {L.}~\bibnamefont
  {Motyka}},\ }\href {\doibase 10.1103/PhysRevLett.102.111802} {\bibfield
  {journal} {\bibinfo  {journal} {Phys.Rev.Lett.}\ }\textbf {\bibinfo {volume}
  {102}},\ \bibinfo {pages} {111802} (\bibinfo {year} {2009}{\natexlab{a}})},\
  \Eprint {http://arxiv.org/abs/0807.2405} {arXiv:0807.2405 [hep-ph]}
  \BibitemShut {NoStop}%
\bibitem [{\citenamefont {Kulesza}\ and\ \citenamefont
  {Motyka}(2009{\natexlab{b}})}]{Kulesza:2009kq}%
  \BibitemOpen
  \bibfield  {author} {\bibinfo {author} {\bibfnamefont {A.}~\bibnamefont
  {Kulesza}}\ and\ \bibinfo {author} {\bibfnamefont {L.}~\bibnamefont
  {Motyka}},\ }\href {\doibase 10.1103/PhysRevD.80.095004} {\bibfield
  {journal} {\bibinfo  {journal} {Phys.Rev.}\ }\textbf {\bibinfo {volume}
  {D80}},\ \bibinfo {pages} {095004} (\bibinfo {year} {2009}{\natexlab{b}})},\
  \Eprint {http://arxiv.org/abs/0905.4749} {arXiv:0905.4749 [hep-ph]}
  \BibitemShut {NoStop}%
\bibitem [{\citenamefont {Beenakker}\ \emph {et~al.}(2009)\citenamefont
  {Beenakker}, \citenamefont {Brensing}, \citenamefont {Kramer}, \citenamefont
  {Kulesza}, \citenamefont {Laenen} \emph {et~al.}}]{Beenakker:2009ha}%
  \BibitemOpen
  \bibfield  {author} {\bibinfo {author} {\bibfnamefont {W.}~\bibnamefont
  {Beenakker}}, \bibinfo {author} {\bibfnamefont {S.}~\bibnamefont {Brensing}},
  \bibinfo {author} {\bibfnamefont {M.}~\bibnamefont {Kramer}}, \bibinfo
  {author} {\bibfnamefont {A.}~\bibnamefont {Kulesza}}, \bibinfo {author}
  {\bibfnamefont {E.}~\bibnamefont {Laenen}},  \emph {et~al.},\ }\href
  {\doibase 10.1088/1126-6708/2009/12/041} {\bibfield  {journal} {\bibinfo
  {journal} {JHEP}\ }\textbf {\bibinfo {volume} {0912}},\ \bibinfo {pages}
  {041} (\bibinfo {year} {2009})},\ \Eprint {http://arxiv.org/abs/0909.4418}
  {arXiv:0909.4418 [hep-ph]} \BibitemShut {NoStop}%
\bibitem [{\citenamefont {Beenakker}\ \emph {et~al.}(2011)\citenamefont
  {Beenakker}, \citenamefont {Brensing}, \citenamefont {Kramer}, \citenamefont
  {Kulesza}, \citenamefont {Laenen} \emph {et~al.}}]{Beenakker:2011fu}%
  \BibitemOpen
  \bibfield  {author} {\bibinfo {author} {\bibfnamefont {W.}~\bibnamefont
  {Beenakker}}, \bibinfo {author} {\bibfnamefont {S.}~\bibnamefont {Brensing}},
  \bibinfo {author} {\bibfnamefont {M.}~\bibnamefont {Kramer}}, \bibinfo
  {author} {\bibfnamefont {A.}~\bibnamefont {Kulesza}}, \bibinfo {author}
  {\bibfnamefont {E.}~\bibnamefont {Laenen}},  \emph {et~al.},\ }\href
  {\doibase 10.1142/S0217751X11053560} {\bibfield  {journal} {\bibinfo
  {journal} {Int.J.Mod.Phys.}\ }\textbf {\bibinfo {volume} {A26}},\ \bibinfo
  {pages} {2637} (\bibinfo {year} {2011})},\ \Eprint
  {http://arxiv.org/abs/1105.1110} {arXiv:1105.1110 [hep-ph]} \BibitemShut
  {NoStop}%
\bibitem [{\citenamefont {Cacciari}\ \emph {et~al.}(2008)\citenamefont
  {Cacciari}, \citenamefont {Salam},\ and\ \citenamefont
  {Soyez}}]{Cacciari:2008gp}%
  \BibitemOpen
  \bibfield  {author} {\bibinfo {author} {\bibfnamefont {M.}~\bibnamefont
  {Cacciari}}, \bibinfo {author} {\bibfnamefont {G.~P.}\ \bibnamefont {Salam}},
  \ and\ \bibinfo {author} {\bibfnamefont {G.}~\bibnamefont {Soyez}},\ }\href
  {\doibase 10.1088/1126-6708/2008/04/063} {\bibfield  {journal} {\bibinfo
  {journal} {JHEP}\ }\textbf {\bibinfo {volume} {0804}},\ \bibinfo {pages}
  {063} (\bibinfo {year} {2008})},\ \Eprint {http://arxiv.org/abs/0802.1189}
  {arXiv:0802.1189 [hep-ph]} \BibitemShut {NoStop}%
\bibitem [{\citenamefont {Cacciari}\ \emph {et~al.}(2012)\citenamefont
  {Cacciari}, \citenamefont {Salam},\ and\ \citenamefont
  {Soyez}}]{Cacciari:2011ma}%
  \BibitemOpen
  \bibfield  {author} {\bibinfo {author} {\bibfnamefont {M.}~\bibnamefont
  {Cacciari}}, \bibinfo {author} {\bibfnamefont {G.~P.}\ \bibnamefont {Salam}},
  \ and\ \bibinfo {author} {\bibfnamefont {G.}~\bibnamefont {Soyez}},\ }\href
  {\doibase 10.1140/epjc/s10052-012-1896-2} {\bibfield  {journal} {\bibinfo
  {journal} {Eur.Phys.J.}\ }\textbf {\bibinfo {volume} {C72}},\ \bibinfo
  {pages} {1896} (\bibinfo {year} {2012})},\ \Eprint
  {http://arxiv.org/abs/1111.6097} {arXiv:1111.6097 [hep-ph]} \BibitemShut
  {NoStop}%
\bibitem [{\citenamefont {Aad}\ \emph {et~al.}(2014{\natexlab{a}})\citenamefont
  {Aad} \emph {et~al.}}]{Aad:2014qaa}%
  \BibitemOpen
  \bibfield  {author} {\bibinfo {author} {\bibfnamefont {G.}~\bibnamefont
  {Aad}} \emph {et~al.} (\bibinfo {collaboration} {ATLAS}),\ }\href {\doibase
  10.1007/JHEP06(2014)124} {\bibfield  {journal} {\bibinfo  {journal} {JHEP}\
  }\textbf {\bibinfo {volume} {1406}},\ \bibinfo {pages} {124} (\bibinfo {year}
  {2014}{\natexlab{a}})},\ \Eprint {http://arxiv.org/abs/1403.4853}
  {arXiv:1403.4853 [hep-ex]} \BibitemShut {NoStop}%
\bibitem [{\citenamefont {Chatrchyan}\ \emph {et~al.}(2014)\citenamefont
  {Chatrchyan} \emph {et~al.}}]{Chatrchyan:2014aea}%
  \BibitemOpen
  \bibfield  {author} {\bibinfo {author} {\bibfnamefont {S.}~\bibnamefont
  {Chatrchyan}} \emph {et~al.} (\bibinfo {collaboration} {CMS}),\ }\href
  {\doibase 10.1103/PhysRevD.90.032006} {\bibfield  {journal} {\bibinfo
  {journal} {Phys.Rev.}\ }\textbf {\bibinfo {volume} {D90}},\ \bibinfo {pages}
  {032006} (\bibinfo {year} {2014})},\ \Eprint {http://arxiv.org/abs/1404.5801}
  {arXiv:1404.5801 [hep-ex]} \BibitemShut {NoStop}%
\bibitem [{\citenamefont {Aad}\ \emph {et~al.}(2014{\natexlab{b}})\citenamefont
  {Aad} \emph {et~al.}}]{Aad:2014wea}%
  \BibitemOpen
  \bibfield  {author} {\bibinfo {author} {\bibfnamefont {G.}~\bibnamefont
  {Aad}} \emph {et~al.} (\bibinfo {collaboration} {ATLAS Collaboration}),\
  }\href@noop {} {\  (\bibinfo {year} {2014}{\natexlab{b}})},\ \Eprint
  {http://arxiv.org/abs/1405.7875} {arXiv:1405.7875 [hep-ex]} \BibitemShut
  {NoStop}%
\bibitem [{\citenamefont {Cowan}\ \emph {et~al.}(2011)\citenamefont {Cowan},
  \citenamefont {Cranmer}, \citenamefont {Gross},\ and\ \citenamefont
  {Vitells}}]{Cowan:2010js}%
  \BibitemOpen
  \bibfield  {author} {\bibinfo {author} {\bibfnamefont {G.}~\bibnamefont
  {Cowan}}, \bibinfo {author} {\bibfnamefont {K.}~\bibnamefont {Cranmer}},
  \bibinfo {author} {\bibfnamefont {E.}~\bibnamefont {Gross}}, \ and\ \bibinfo
  {author} {\bibfnamefont {O.}~\bibnamefont {Vitells}},\ }\href {\doibase
  10.1140/epjc/s10052-011-1554-0, 10.1140/epjc/s10052-013-2501-z} {\bibfield
  {journal} {\bibinfo  {journal} {Eur.Phys.J.}\ }\textbf {\bibinfo {volume}
  {C71}},\ \bibinfo {pages} {1554} (\bibinfo {year} {2011})},\ \Eprint
  {http://arxiv.org/abs/1007.1727} {arXiv:1007.1727 [physics.data-an]}
  \BibitemShut {NoStop}%
\bibitem [{\citenamefont {Read}(2002)}]{Read:2002hq}%
  \BibitemOpen
  \bibfield  {author} {\bibinfo {author} {\bibfnamefont {A.~L.}\ \bibnamefont
  {Read}},\ }\href {\doibase 10.1088/0954-3899/28/10/313} {\bibfield  {journal}
  {\bibinfo  {journal} {J.Phys.}\ }\textbf {\bibinfo {volume} {G28}},\ \bibinfo
  {pages} {2693} (\bibinfo {year} {2002})}\BibitemShut {NoStop}%
\end{thebibliography}%
\end{document}